\documentstyle[sprocl]{article}

\input{psfig.sty}

\def\PLB{{\em Phys. Lett.}  B}
\def\PRL{\em Phys. Rev. Lett.}
\def\PRD{{\em Phys. Rev.} D}

\def\be{\begin{equation}}
\def\ee{\end{equation}}
\def\bea{\begin{eqnarray}}
\def\eea{\end{eqnarray}}
\def\eslt{E\llap/_T}
\def\to{\rightarrow}
\def\Re{{\cal R \mskip-4mu \lower.1ex \hbox{\it e}}\,}
\def\Im{{\cal I \mskip-5mu \lower.1ex \hbox{\it m}}\,}

\def\te{\tilde e}

\def\tG{\tilde G}
\def\tb{\tilde b}
\def\tt{\tilde t}

\def\ttau{\tilde \tau}
\def\tg{\tilde g}

\def\tnu{\tilde\nu}
\def\tmu{\tilde\mu}
\def\ttau{\tilde\tau}
\def\tell{\tilde\ell}
\def\tq{\tilde q}

\def\tw{\widetilde W}
\def\tz{\widetilde Z}
\def\agt{\stackrel{>}{\sim}}

\begin{document}

\hfill{UH-511-911-98}
\vspace{5mm}
\title{DEVELOPMENTS IN SUPERSYMMETRY PHENOMENOLOGY~\footnote{Plenary
talk presented at the Sixth International Symposium on Particles,
Strings and Cosmology, PASCOS-98, North Eastern University, Boston, MA
02146, March, 1998.}}

\author{XERXES TATA}

\address{Department of Physics and Astronomy,\\ 
University of Hawaii, Honolulu, HI 96822, USA } 


\maketitle\abstracts{We survey strategies generally employed for SUSY
discovery at colliders and then discuss how these may have to be altered
for SUSY searches at the Tevatron if $\tan\beta$ is large. We also
discuss the reach of the Tevatron and the LHC in
gauge-mediated SUSY breaking scenarios, assuming that the NLSP decays
into photons. Finally, we briefly recapitulate measurements (which serve
to guide us to the underlying theory) that might
be possible at future colliders if supersymmetry is discovered.}

Weak scale supersymmetry is now a mature subject, and much theoretical
and experimental effort has been expended in the development of
strategies for sparticle searches at colliders. As yet no direct signal
has been found in experiments at LEP2 and at the Tevatron. The absence
of any signal has been translated into lower limits on sparticle
masses. This translation is always done within the framework of a model,
frequently taken to be the mSUGRA model, or sometimes, the MSSM
supplemented with {\it ad hoc} assumptions about squark/slepton masses
and gaugino masses, together with the conservation of $R$-parity. Some
searches are less sensitive to these assumptions than others: for
example, it is not unreasonable for charged slepton searches at LEP to
assume that other charged sparticles are heavier (since they would
otherwise have been found), and that the slepton
directly decays to the lightest SUSY particle (LSP) which is frequently
the neutralino $\tz_1$. For other searches, such as those for squarks
and gluinos at the Tevatron or LHC, cascade decays are very important,
and
the signals depend on properties of all sparticles lighter than
these.

Analyses from LEP2 have resulted in bounds~\cite{MNICH} $m_{\tw_1}
\agt 87-91$~GeV, $m_{\tell_R} \agt 85$~GeV and $m_{\tmu_R} \agt 75$~GeV,
the exact value depending on $m_{\tz_1}$. At the Tevatron, the D0 and
CDF collaborations have reported negative results for their searches for
gluinos and squarks in the $\eslt$ as well as dilepton channels (the CDF
collaboration has searched in the same sign (SS) dilepton channel
characteristic of the production of Majorana gluinos), resulting in a
lower mass bound of $m_{\tg} \agt 180$~GeV ($\agt 250$~GeV if $m_{\tq}
\simeq m_{\tg}$). These limits, which are reviewed in Ref.~\cite{CCEFM},
vary somewhat with the model, and may be considerably weaker for some
ranges of mSUGRA parameters. It is significant that the bounds
via the dilepton channels are already competetive with those from the
$\eslt$ channel. With a data sample of several $fb^{-1}$ that will be
accumulated at the Main Injector (MI) or its luminosity upgrades, the
reach via multilepton channels may exceed that via the canonical $\eslt$
search.  Within any framework, such as mSUGRA, where
electroweak gauginos are expected to be considerably lighter than the
coloured gluinos, essentially background-free clean trilepton events (that are
free from hadronic activity other than that due to QCD radiation) from
$\tw_1\tz_2$ production potentially offer the greatest reach for
supersymmetry at the Tevatron: for values of parameters where leptonic
decays of $\tw_1$ and $\tz_2$ are strongly enhanced, experiments in the
MI era may probe $m_{1/2}$ values as large as 225~GeV, corresponding to
gluinos as heavy as 600~GeV. It should, however, be kept in mind that
there are other regions of parameter space where the branching fraction
for the decay $\tz_2 \to \ell \bar{\ell} \tz_1$ is strongly suppressed,
so that {\it there may be no detectable trilepton signal even for charginos
below the LEP bound.} Thus, while the clean trilepton channel may lead to
a spectacular discovery of SUSY, non-observation of any signal will not
result in unambiguous lower bounds of $m_{\tw_1}$ or $m_{\tz_2}$.

Our discussion up to now has been limited to small values of the
parameter $\tan\beta$ for which bottom and $\tau$ Yukawa couplings may
be neglected. For larger values of $\tan\beta$, these Yukawa
interactions have two important effects.\cite{DREES} First, they lead to
a reduction of $m_{\tb_1}$ and $m_{\ttau_1}$ relative to the masses of
corresponding first and second generation sfermions.
Second, when Yukawa couplings are sizeable,
contributions to chargino and neutralino decays mediated by Higgs scalars
as well as those due to Higgsino components of neutralinos and
charginos can considerably modify chargino and especially neutralino
decay patterns. Both these effects enhance decays of neutralinos,
charginos and even gluinos to third generation particles as illustrated in
Fig.~1. 
\begin{figure}
\centerline{\psfig{file=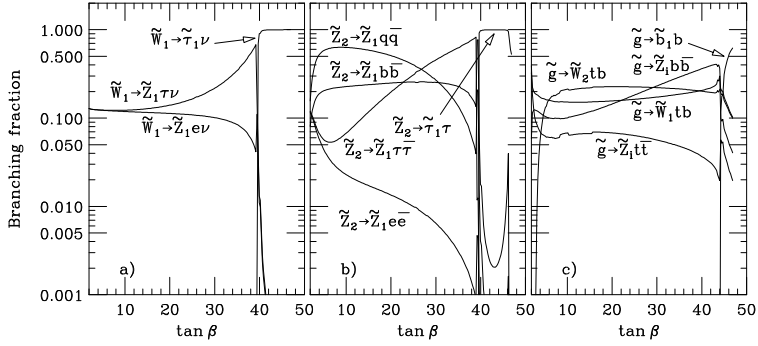,height=4.5cm,angle=90}}
\caption[]{The dependence of various branching fractions on
$\tan\beta$. In {\it a}) and {\it b}), we take the parameters
($m_0,m_{1/2},A_0)=(150,150,0)$ GeV while {\it c}) uses
($m_0,m_{1/2},A_0)=(700,250,0)$ GeV. In all frames, $\mu >0$ and
$m_t=170$ GeV.}
\end{figure}
We see that Yukawa coupling effects show up as non-universal
leptonic branching fractions of $\tw_1$ and $\tz_2$ for $\tan\beta$ as
small as 5-10. For still larger values of $\tan\beta$, the decays to tau
(bottom) increasingly dominate chargino and neutralino (gluino) decays,
and for extremely large values of $\tan\beta$ two body decays to third
generation sfermions may become accessible. 

These modifications due to Yukawa couplings have been incorporated
in the simulation program ISAJET.~\cite{ISAJET} The most important effect of these is
that they reduce the cross sections for clean or jetty isolated
multilepton events which we have just seen potentially provide the
greatest SUSY reach at the luminosity upgrades of the Tevatron. In order
to assess the prospects for detecting supersymmetry signals with tagged
$b$-jets or tau leptons, we have performed an ISAJET simulation where in
addition to the usually searched for ({\it a})~$\eslt$ + jets, ({\it
b})~multijets + $n$~lepton ($e$ or $\mu$), $n=1-3$ (SS and opposite sign (OS)
for $n=2$), ({\it c})~clean OS dileptons and ({\it d})~clean
trilepton channels, we have also examined the detectability of SUSY at
the Tevatron via ({\it e})~multijet + tagged $b$/$\tau$, ({\it f})
multijet + OS and SS dilepton or trilepton
channels where at least one of the leptons is identified as $\tau$, and
finally, ({\it g})~clean OS dilepton and trilepton channels where one of
the leptons is identified as tau. In our simulation, we assume that jets
with $E_T> 15$~GeV
with just 1 or 3 charged prongs within 10$^o$ of the axis and no other
charged prongs within 30$^o$ and with invariant mass smaller than
$m_{\tau}$ are
hadronically decaying taus. We also identify central $b$ jets with an
efficiency of 50\%. The result of our computation\cite{DREES} is summarized in
Fig.~2 where we have scanned the $m_0-m_{1/2}$ plane for several values
of $\tan\beta$ to see if there is an observable signal in {\it any} of
these channels. We have taken $A_0=0$. 
The grey (hollow) squares denote points where there is a
5$\sigma$ signal above background with a minimum of 5 expected signal
events, assuming an integrated luminosity of 2~$fb^{-1}$ (25~$fb^{-1}$)
at a $\sqrt{s}=2$~TeV $p\bar{p}$ collider. The bricked (shaded) regions
of the plane are excluded by theoretical (experimental) constraints. 
Several comments are in order.
\begin{itemize} 

\item The $\eslt$, $\eslt + b$, clean $3\ell$ and clean $3\ell$ with an
identified tau channels establish the entire plot, though for some
points a signal may also be observable in other channels. The region
where the signal should be observable falls sharply with increasing
values of $\tan\beta$. Indeed for the largest values of $\tan\beta$ we
see that there are no regions beyond LEP bounds where there will be an
experimental signature at the MI.
\begin{figure}
\centerline{\psfig{file=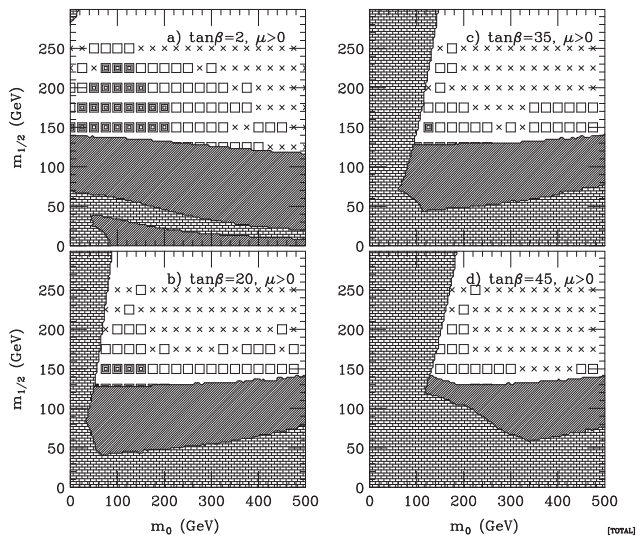,height=7cm,angle=0}}
\caption[]{The combined reach of the Tevatron MI (2~$fb^{-1}$) and TeV33
(25~$fb^{-1}$) for 
mSUGRA via {\it all} of the signal channels considered here.}
\label{fig2}
\end{figure}
\item We have used rather hard cuts for the clean lepton channels which
allow secondary $e$ and $\mu$ from decays of taus to pass with a very
low efficiency. A subsequent study by the Wisconsin Group~\cite {WISC}
has shown that by using softer cuts for the clean trilepton sample,
there would be additional ranges of parameters where
a signal is just observable above physics backgrounds even if
$\tan\beta$ is large. Potentially worrisome
non-physics backgrounds where a soft jet is mis-identified as one of the
leptons, or a heavy quark fakes an isolated lepton, are thought to
be~\cite{KAMON} under control.
  
\item Although the cross section for events with hadronically decaying
taus is large, such taus are identified with very low efficiency in our
simulation because we require the visible hadronic decay products to have 
$E_T > 15$~GeV. We urge our experimental colleagues to examine
strategies to reliably identify hadronically decaying taus with smaller
visible $E_T$, since this could significantly extend the region of the
mSUGRA plane where there would be an observable signal if $\tan\beta$
happens to be large. This is desirable even in light of the soft lepton
signal mentioned above, since observation of a tau excess would indicate
a source of lepton non-universality (though not necessarily large
$\tan\beta$).

\end{itemize}

Within the mSUGRA framework, $\tg\tg$, $\tg\tq$ and $\tq\tq$ production
are the dominant SUSY processes at the LHC if gluinos and squarks are
lighter than 1-2~TeV. Once produced, the gluinos and squarks cascade
decay to charginos and neutralinos, until the cascade terminates in the
stable LSP. Sparticle production is thus signalled by multijet +
multilepton + $\eslt$ events. There may also be clean multilepton events
from $\tw_1$ and $\tz_2$ production, but these yield a smaller
reach. Several studies~\cite{CMS,BAERLHC} have shown that within this
framework, experiments at the LHC should be able to probe gluinos and
squarks even if they are as heavy as 2~TeV. Although these studies have
been performed for low $\tan\beta$, unlike for the Tevatron, the LHC
reach is more or less independent of $\tan\beta$. The reason is that for
a large range of mSUGRA parameters, $m_{\tz_1} \simeq
\frac{1}{2}m_{\tw_1} \simeq \frac{1}{2}m_{\tz_2} \simeq \frac{1}{6}
m_{\tg}$, so that charginos and neutralinos have two body decays to real
$W$, $Z$ and Higgs bosons if gluinos are heavy; their branching
fractions to third generation fermions are thus determined by the decays
of the bosons, and so are insensitive to $\tan\beta$, except for small
values of $m_0$ where decays to $\ttau_1$ may be accesible but not
decays to other sleptons. Our analysis~\cite{PREP} bears out that
for large $\tan\beta$ the region of the $m_0-m_{1/2}$ plane that can be
explored at the LHC is qualitatively similar to that in our previous
studies,~\cite{BAERLHC} except for low values of $m_0$ where the reach
in $m_{1/2}$ is reduced from over 1~TeV to about 850~GeV.
For parameter ranges where chargino and neutralino decays to
stau are accessible or where three body decays dominate, we may expect
significant lepton non-universality in SUSY events. We are currently
exploring whether a detailed study of the tau rate and polarization
could provide an independent indication that $\tan\beta$ is large.

Supersymmetry should be readily discoverable~\cite{MUR} at future $e^+e^-$
colliders, provided these have sufficent energy to pair produce charged
sparticles or visibly decaying sneutrinos. 
The projected reach of Linear Colliders is compared to that of the LHC
and Tevatron upgrades in Fig.~3, where the bricked and hatched regions
have the same meaning as before.
\begin{figure}
\centerline{\psfig{file=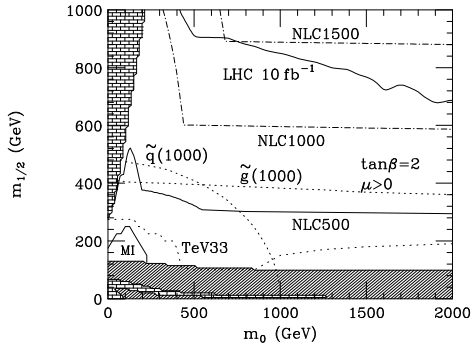,height=4.5cm,angle=90}}
\caption[bth]{The SUSY reach for various facilities as given by the mSUGRA
model. We take $\tan\beta=2$, $A_0=0$ and $\mu > 0$.}
\label{fig:SUGRA}
\end{figure}
For the purposes of reach and for this purpose alone, a Linear Collider
with a centre of mass energy between 1 and 1.5~TeV would have an mSUGRA
reach comparable to the LHC. 

Sparticle masses can be precisely (1-2\% for $\tell,\tnu, \tw_1$ and
$\sim$5\% for $\tt_1$) measured~\cite{MUR} at Linear Colliders, even if
these cascade decay to the LSP.  If SUSY is discovered, it should be
possible to set up an essentially model-independent
program~\cite{SNOWNLC} at the NLC to measure sparticle properties. Such
measurements provide us with guide posts that may help us unravel the
dynamics of SUSY breaking. The availability of a polarized electron beam
as well as variable but precisely tuned beam energy are crucial for
these measurements.  Some mass measurements,~\cite{HINCH} particularly
of mass differences from a determination of kinematic endpoints, are
also possible at the LHC.  Indeed, it has been shown that it should be
possible to use LHC measurements to test for consistency of (and hence
falsify!) well-defined models, such as mSUGRA or gauge-mediated SUSY
breaking models, with relatively few parameters. Various case studies
show that it should also be possible to determine the mSUGRA
parameters $m_{1/2}$ and $m_0$ from the data.
It is, however, presently unclear how to directly go from LHC data to
the underlying model. This is important since it is quite possible that
none of the models that we have thought about will turn out to be
right. In fairness though, such studies have only just begun, and much
work remains to be done along this direction.

We have become accustomed to  various {\it assumptions} about
symmetries of 
physics at very high energy, that leads to the mSUGRA model (with
its universal parameters) as the
effective theory below the Planck scale. It is important to keep in mind
that these assumptions may prove to be invalid. Testing these
assumptions should be an integral part of any supersymmetry program, and
our discussion shows that this should be possible at
future $pp$ and lepton supercolliders. 

Many SUSY searches (especially at hadron colliders) rely on $\eslt$ from
the undetected LSPs to discriminate the SUSY signal from Standard Model
backgrounds. In models where $R$-parity is not conserved, the LSP
(frequently the lightest neutralino) decays into ordinary particles, and
the $\eslt$ is considerably reduced. Assuming a minimal particle
content, at least one of lepton or baryon numbers is not conserved in
$R$-parity violating models. In the former case, the LSP decays into
leptons (possibly together with jets), so that even though $\eslt$ is
reduced, SUSY events result in increased cross sections for isolated
multilepton topologies which have small backgrounds. In the most
favourable case where the LSP decays only via $\tz_1 \to
\ell\bar{\ell'}\nu$ ($\ell,\ell'=e,\mu$), it should be possible to explore
parameter ranges corresponding to $m_{\tg}=800$~GeV even at the
MI.~\cite{TEVRPV} The (almost) worst case scenario, however, is when the
LSP decays entirely hadronically via $\tz_1 \to qqq$ (or
$\bar{q}\bar{q}\bar{q}$): not only is the $\eslt$ reduced, but leptons
from chargino and neutralino decays find it harder to remain
isolated. There may then be no observable signal~\cite{TEVRPV} at the MI
if gluinos are heavier than about 200~GeV. Assuming, however, that LSP decay is
the only significant effect of $R$ violating interactions, it has been
shown~\cite{RVIOL} that experiments at the LHC should still be able to
probe gluino and squarks with masses up to 1~TeV even in this
unfavourable case.

Models where gauge interactions, and not gravity, serve to mediate SUSY
breaking effects to Standard Model particles and their superpartners
have recently received a lot of attention. In these scenarios, the SUSY
breaking scale $\sqrt{F}$ can be (but is not necessarily) as low as
${\cal O}(100)$~TeV, so that the gravitino mass $m_{\tG} \sim F/M_P$ can
be as small as ${\cal O}$(1~eV). The coupling of the longitudinal components
of such superlight gravitinos to 100~GeV sparticles, while small
compared to gauge couplings, is not negligible since it may cause the
next lightest SUSY particle (NLSP) to decay into the gravitino within
the experimental apparatus. The decay patterns of other sparticles are
still governed by their (much larger) gauge and Yukawa couplings, and
are essentially unaffected.

Collider signatures from such models depend very much on what the NLSP
is, and also on its lifetime.  If the NLSP is a neutralino, it can decay
into a photon (and possibly also a $Z$ or Higgs boson) and a
gravitino. If it decays outside the detector, the expected topologies
are qualitatively similar to those within the mSUGRA model, though the
relative cross sections may be different on account of differences in
the mass spectra. If the NLSP decays inside the detector, SUSY events
contain up to two isolated photons in addition to jets and leptons. The
unstable NLSP need not be neutral. In many models, it is a charged
slepton (frequently stau), in which case charginos and neutralinos
(which may be secondary products of squark and gluino decays) decay into
the stau/slepton ($\te_R,\tmu_R$) NLSP which then decays into tau/($e$
or $\mu$) and a gravitino. SUSY events then contain an abundance of
taus/leptons. The NLSP may be quite long lived, so
that its decay products emerge from a secondary vertex whose
displacement could provide an additional handle on the signal:~\cite{GUN}
in particular, it could yield a measure of the SUSY breaking
scale. A very long-lived charged NLSP that makes it through the detector
would be signalled~\cite{FENG} by highly ionizing tracks.

We have simulated signals from gauge mediated models with a neutralino
NLSP that decays promptly into photons using ISAJET. In the simplest of
these models, Tevatron experiments would already be
sensitive~\cite{GMLESB} to values of the parameter $\Lambda$ (which
fixes the sparticle mass scale) smaller than 50-60~TeV (corresponding to
$m_{\tg}=450$~GeV) via the multijet + isolated photons channel, while
the corresponding reach of the MI would be as high as
$m_{\tg}=800$~GeV. The LHC should, via a search for inclusive
$\gamma\gamma + jets +\eslt$ events, be able to explore~\cite{PEDRO}
$\Lambda \leq 400$~TeV; this corresponds to $m_{\tg} \sim 2.8$~TeV, for
which chargino and neutralino production is the dominant SUSY
process. Finally we note that the $E_T$ spectrum of the photon in these
events scales with the NLSP mass, and may yield information about the
underlying parameters of the model. A study of this issue as well as of the
reach of the LHC when the NLSP is a slepton is currently in
progress.

\section*{Acknowledgments}
I am grateful to H. Baer, M.~Brhlik, C. Chen, M. Drees, C. Kao,
P. Mercadante, F. Paige and Y. Wang for collaboration on the work
reported here.  This research was supported, in part, by U.S. DoE grant
no. DE-FG-94ER40833.

\end{document}